\documentclass[twocolumn,floatfix,superscriptaddress,amsmath,showpacs,showkeys,aps,prb]{revtex4}

\usepackage{t1enc}
\usepackage{times}

\begin{document}

\bibliographystyle{prsty}

\title{Orientational order in two dimensions from competing
  interactions at different scales}
\author{Daniel G.\ Barci}
\affiliation{Departamento de F{\'\i}sica Te\'orica,
Universidade do Estado do Rio de Janeiro, Rua S\~ao Francisco Xavier 524, 20550-013,  Rio de Janeiro, RJ, Brazil.}
\altaffiliation{Research Associate of the Abdus Salam International Centre for Theoretical Physics}
\author{Daniel A.\ Stariolo}
\affiliation{Departamento de F{\'\i}sica, Universidade Federal do Rio
  Grande do Sul, CP 15051, 91501-970, Porto Alegre, Brazil} 
\altaffiliation{Research Associate of the Abdus Salam International Centre for Theoretical Physics}
%
\begin{abstract}
We discuss orientational order in two dimensions in the context of
systems with competing isotropic interactions at different scales.
We consider an extension of the Brazovskii model for stripe phases
including  explicitly quartic terms with nematic symmetry in the energy.
We show that leading fluctuations of the mean field nematic solution
drive the isotropic-nematic transition into the Kosterlitz-Thouless universality class, i.e. these
systems have a thermodynamic phase with orientational quasi-long-range order.
\end{abstract}
%
%
%
%

\maketitle

\section{Introduction}
Systems with competing interactions at different length scales are
common in nature~\cite{SeAn1995}. 
Examples go from highly correlated quantum systems like quantum Hall
samples~\cite{FrKi1999} 
and high 
$T_c$ superconductors~\cite{KiFrEm1998}, to classical systems like
ferromagnetic ultrathin films~\cite{AbKaPoSa1995,VaStMaPiPoPe2000}, 
diblock copolymers~\cite{FrHe1987,BaRoFr1990}, colloidal
suspensions~\cite{ImRe2006},
ferromagnetic garnet
films~\cite{SeMoGoWo1991}
 and magnetic fluids~\cite{DiErGoJaLa1993}.
 The essential phenomenology of these kind of
systems was described in a classic paper by
Brazovskii~\cite{Br1975}. Competition on different scales gives
rise to ordered phases dominated by a non-zero wave vector in
reciprocal space, as opposed to the usual $k=0$ long range order. The
non-zero value $k_0$ of the dominant wave vector gives rise to
spatially modulated structures. In three dimensions Brazovskii showed
that
 striped phases appear through a first order phase
transition induced by fluctuations~\cite{Br1975}. Stripe patterns
 show both positional (anisotropic) and orientational
long range order, although the stripe solutions in the self consistent
Hartree approximation are marginally stable in three dimensions.
Subsequent works extended the original results to two dimensional
systems in spite of the fact that, strictly speaking, fluctuations
prevent any long range order~\cite{SwHo1977,HoSw1995}.

Motivated by recent experimental observations of phases with modulated
order in two dimensional systems, we analyzed in a recent
letter~\cite{BaSt2007} the conditions for the existence of a purely
orientational phase, a nematic phase, in models of the Brazovskii
class. In the framework of the Renormalization Group, we showed that an isotropic-nematic phase transition is
generically present in these kind of models, provided suitable quartic interactions
between the basic degrees of freedom are taken into account. These interactions  are naturally generated in
the renormalization process. Furthermore, we found that, in two
dimensions,  renormalization of the Brazovskii model gives rise to an
infinite number of relevant terms which makes the model
non-renormalizable. We have shown in Ref.\ \onlinecite{BaSt2007} that
all those terms possess a common symmetry under rotations by $\pi$, a nematic symmetry. 
Keeping only the term with the
highest symmetry, corresponding to quadrupole-quadrupole interactions,  
we showed that a isotropic-nematic phase transition
is present and that it is of second order at mean field level.  
Nevertheless, it was anticipated that
the nature of the transition would probably be affected upon inclusion of
fluctuations~\cite{BaSt2007}, since it is not possible to break a continuous symmetry 
in two dimensions with short ranged interactions~\cite{MeWa1966}.

The original Brazovskii model in
three dimensions is at its lower critical dimension, 
with fluctuations in the stripe solutions diverging
logarithmically with the linear size of the system. The
situation is more delicate in two dimensions where fluctuations are  linearly divergent~\cite{SwHo1977}. 
Then,
if some order of this kind survives in two dimensions it must be
purely orientational. However, it is still necessary to check whether
orientational order survives to fluctuations
of the relevant order parameter. Of course, in real systems, the nematic
or even the stripes phase can be stabilized by other factors, like
anisotropies coming from the lattice substrate, the presence of
impurities or some disorder that pin the stripe order. These effects
will not be considered here, where we rest at the level of a completely
isotropic system.

In the present work we  pursue the analysis of a generalized Brazovskii model which takes into account 
quadrupolar interactions in two dimensions. We briefly review the mean field treatment of  Ref.\ \onlinecite{BaSt2007}, 
and we evaluate thermal fluctuations. We show that the isotropic-nematic transition belongs to the well known Kosterlitz-
Thouless universality class~\cite{KoTh1973}, i.e. upon inclusion of order parameter
fluctuations, the mean field solution with nematic long range order
in fact retains only quasi-long-range orientational order. 
In turn,
the stripe solution is unstable to fluctuations and a possible
smectic-like phase reduces to a point at zero temperature. 
Similar results were found long time
ago by Toner and Nelson in the context of defect mediated
melting in two dimensions  ~\cite{ToNe1981}. In fact, both approaches
are complementary and consistently lead to the same phase diagram.
The present results were briefly anticipated by us
 in a reply ~\cite{BaSt2007-2} to a comment\cite{Le2007} to Ref.\ 
 \onlinecite{BaSt2007}. 

In the following, we introduce in \S \ref{LG} a prototypical model of the
Landau-Ginzburg type for a nematic order parameter. We show that,
while the mean field treatment leads to 
 a  second order phase transition in two dimensions, low energy
 fluctuations diverge logarithmically, as in the XY model of
magnetism, destroying the long range order and leading to an algebraic
decay of the correlations. 
In \S \ref{Braz}  we
introduce the extended Brazovskii model considered in
Ref.\ ~\onlinecite{BaSt2007}, and briefly discuss the mean field solution. We show that the relevant
leading
order fluctuations can be mapped to an XY type model.
We thus show that, within the Gaussian approximation, 
 the isotropic-nematic phase transition is in the Kosterlitz-Thouless
universality class. The analysis also allows us to express
the effective elastic constant $K(T)$ as a function of the parameters of the
original model. A brief discussion of the results is given in \S \ref{Conclusions}.

\section{Landau-Ginsburg theory for the nematic transition\label{LG}}
The Landau-Ginzburg theory for a three dimensional nematic phase is
well known~\cite{deGPr1998}. Here, we briefly restate it for a two
dimensional system because it presents some characteristics exclusive of the
dimensionality of the problem. This analysis will be also helpful as a guide to the evaluation  
of  fluctuations of the specific model of section \ref{Braz}.

\subsection{Order parameter}
\label{op}

The order parameter for a $2d$ nematic has two components to
identify an orientation in the plane 
and the intensity, but not a direction. 
This means that it must be symmetric under the transformation $\theta\to \theta+\pi$. 

Consider  a complex number written in the form:
\begin{equation}
Q=\alpha \; e^{i 2\theta}
\label{Qcomplex}
\end{equation}
where $\theta$ is an angle in $2d$ space.  Then, if $<Q>\neq 0$, the
phase is said to have 
orientational order in the $\theta$ direction with the nematic symmetry $\theta\to \theta+\pi$. 
The nematic symmetry implies that the order parameter is not a
vector. Instead, we can arrange the real and imaginary parts of (\ref{Qcomplex})
in a second rank symmetric 
and traceless tensor in the the following way:
\begin{eqnarray}
{\cal R}e(Q)&=&\alpha\; \cos(2\theta)\equiv Q_{xx}=-Q_{yy} \nonumber \\ 
{\cal I}m(Q)&=&\alpha\; \sin(2\theta)\equiv Q_{xy}=Q_{yx} 
\end{eqnarray}
 and
\begin{equation}
\hat Q=  \left(
\begin{array}{cc}
Q_{xx} &  Q_{xy} \\
Q_{xy} & -Q_{xx}
\end{array}
\right)
=\alpha \left(
\begin{array}{cc}
\cos(2\theta) & \sin(2\theta) \\
\sin(2\theta) & -\cos(2\theta)
\end{array}
\right)
\label{Qtensor}
\end{equation}
Defining now a unit vector (the director) $\hat n$ with components $n_x=\cos\theta$
and $n_y=\sin\theta$, the nematic order parameter reads:
\begin{equation}
\hat Q  
=\alpha \left(
\begin{array}{cc}
n_x^2-n_y^2 & 2 n_x n_y \\
2 n_x n_y & n_y^2-n_x^2
\end{array}
\right),
\label{Qnx}
\end{equation}
or in component notation:
\begin{equation}
\hat Q_{ij}  
=2 \alpha \left( n_i \,n_j-\frac{1}{2}n^2\delta_{ij} 
\right)
\label{Qn}
\end{equation}
Equations (\ref{Qcomplex}) and (\ref{Qn}) are two different ways of writing the same thing. 
We can now developed a Landau-Ginzburg free energy for a constant tensor near the
transition, where $<Q>$ is very small. The leading rotational invariant terms are:
\begin{equation}
F(\hat Q)= \frac{1}{4} a_2 Tr(\hat Q^2) + \frac{1}{8} a_4 Tr(\hat Q^4)+\ldots 
\label{Ftensor}
\end{equation}

Using Eq. (\ref{Qtensor}) it is very simple to show that the free
energy reduces to:
\begin{equation}
F(\alpha)= \frac{1}{2}a_2 \alpha^2 +\frac{1}{4} a_4 \alpha^4+\ldots 
\label{Falpha}
\end{equation}
Note that  at this level the free energy is independent of $\theta$,
 that means that it is invariant under arbitrary global rotations.
In particular, in two dimensions, $Tr(\hat Q^3)=0$ and therefore there are no terms with $\alpha^3$, at
 variance with the $3d$ case. This implies that the mean field
 isotropic-nematic transition is of second order in $2d$.

\subsection{Mean field phase transition}

Consider the free energy of Eq. (\ref{Falpha}) and suppose that
$a_4>0$. Therefore, if $a_2>0$ the only 
minimum of this energy is $\alpha=0$ and then $<Q>=0$. Conversely, if
$a_2<0$ the minimum is 
at $\alpha=(-a_2/a_4)^{1/2}$ and 
\begin{equation}
<Q> = \sqrt{\frac{-a_2}{a_4}}\; e^{i 2 \theta} ,
\end{equation}
or in terms of the director components:
\begin{equation}
<\hat Q_{ij}>  
 = 2 \sqrt{\frac{-a_2}{a_4}}\; \left( n_i  n_j-\frac{1}{2}n^2\delta_{ij} 
\right)
\label{Q}
\end{equation}
To leading order (near the transition) $a_2(T)=a\; (T-T^*)$, where $a>0$
is a constant and $T^*$
is the critical temperature. At the critical point the rotational
symmetry in the plane is spontaneously broken. Choosing the director
direction to correspond to $\theta=0$ then:
\begin{equation}
\langle Q\rangle =\left\{
\begin{array}{lcl}
0 & \mbox{if} & T>T^*   \\
 & &  \\
 \sqrt{\frac{a}{a_4}} \; \left(T^*-T\right)^{1/2}
 & \mbox{if}&  T<T^*
\end{array}
\right.
\end{equation}
This is the classic Landau-Ginzburg scenario for a second order phase transition with $T_c=T^*$. We
will see that fluctuations in the director orientation change this picture. 

\subsection{Fluctuations}

We have developed the free energy of Eq.(\ref{Falpha}) considering
that the order parameter $Q$ is constant. 
However, if we want to study local fluctuations we can consider a
local order parameter of the form $Q\equiv Q(x)$ 
and study the free energy for small variations of $Q(x)$ around the
mean field value $Q$. In order to do this we
need to introduce terms proportional to derivatives of the order
parameter
in the expansion of 
the free energy . 
To leading order, we consider just first derivatives of $Q$ and write a rotational invariant free energy of the form:
\begin{eqnarray}
F(\hat Q)&=& \frac{1}{V}\int d^2x\;\left\{ \frac{\rho}{4}  Tr(\hat Q \hat D \hat
Q ) \right. \nonumber \\
&+& \left. \frac{1}{4} a_2 Tr(\hat Q^2) + 
\frac{1}{8} a_4 Tr(\hat Q^4)+\ldots \right\}
\label{Ftensorinhom}
\end{eqnarray}
where $\rho$ is a stiffness constant and the symmetric derivative tensor $\hat D_{ij}\equiv \nabla_i\nabla_j$. 

Because the free energy is symmetric under global rotations, 
low energy angle fluctuations are the most relevant modes that rule
the behaviour of the system. 
We will see that in $2d$ the angular correlations are logarithmically
divergent, ruling out true long range order but showing instead
quasi-long-range order or power-law decay of spatial correlations. 
Consider a local order parameter of the form
\begin{equation}
Q(x)=\sqrt{\frac{-a_2}{a_4}}\; e^{i 2 \theta(x)} 
\label{GMcomplex}
\end{equation}
or in tensor form, as a function of the director components:
\begin{equation}
\hat Q_{ij}(x) =2 \sqrt{\frac{-a_2}{a_4}}\; \left( n_i(x) n_j(x)-\frac{1}{2}n^2\delta_{ij} 
\right)
\label{GMn}
\end{equation}

Thus, fixing the modulus to its mean field value, we proceed to study
small local fluctuations in the direction of the director 
in the nematic phase. 
Replacing Eq.(\ref{GMcomplex}) or (\ref{GMn}) into
(\ref{Ftensorinhom}) we find
that:
\begin{equation}
\delta F\equiv F(Q(x))-F(<Q>)= K(T)\; \int d^2x\; |\vec\nabla
\theta(x)|^2,
\label{deltaf}
\end{equation}
where 
\begin{equation}
K(T)= \frac{2|a_2|}{a_4}\rho.
\end{equation}
Therefore, the free energy for the small angle fluctuations of the
director can be mapped into the free energy of the XY
model~\cite{KoTh1973}. 
Angle correlations in the XY model decay algebraically as:
\begin{equation}
\langle \cos{(\theta(x)-\theta(0))} \rangle \propto x^{-\eta}
\end{equation}
with $\eta=T/2\pi \rho$.
Then, the isotropic-nematic transition in $2d$ belongs to the Kosterlitz-Thouless
universality class with a disordering mechanism mediated by the
unbinding of topological defects~\cite{KoTh1973}. The only difference is that the
role of vortices in the XY model is played here by disclinations~\cite{ToNe1981}.

These results are independent of any microscopic mechanism. In the
present case, if we begin with a Brazovskii 
type Hamiltonian, one should be able to reach Eq. (\ref{Ftensorinhom})
where  the parameters 
$a_2$, $a_4$, $\rho$ and $T^*$ should be written in terms of the more
``microscopic'' ones\cite{BaSt2007}. 
This is the subject of the next sections. 

\section{Model with competing isotropic interactions at different scales \label{Braz}}

Long time ago Brazovskii~\cite{Br1975} introduced a rather general model with
the aim of
capturing the physics of systems with isotropic competing interactions
at different scales. The model should be relevant for a wide class of
systems as discussed in the Introduction.
Specializing to two spatial dimensions and considering a
scalar order parameter (Ising symmetry ), the Brazovskii model is
defined (in reciprocal space) by a 
coarse-grained Hamiltonian of the type:
\begin{equation}
H_0= \int_{\Lambda} \frac{d^{2}k}{(2\pi)^2}\;\phi(\vec k)\left(r_0 + \frac{1}{m}(k-k_0)^2+ \ldots\right) \phi(-\vec k)
\label{H0}
\end{equation}
where $r_0(T)\sim a(T-T_c)$, $k=|\vec k|$ and $k_0=|\vec k_0|$ is a constant given by the nature of the competing 
interactions. $\int_{\Lambda} d^{2}k \equiv  \int_0^{2\pi} d\theta \int_{k_0-\Lambda}^{k_0+\Lambda} dk\;k$
and $\Lambda \sim \sqrt{m r_0}$ is a cut-off where the expansion of the free energy up to quadratic order in the momentum makes 
sense. The ``mass'' $m$ measures the curvature of the dispersion relation around the minimum $k_0$ and  the ellipsis 
in eq.(\ref{H0}) indicates higher order terms in $(k-k_0)$. 
The correlator has a maximum at $k=k_0$ with a correlation length $\xi\sim 1/\sqrt{m r_0}$. Therefore, near 
criticality ($r_0\to 0$), the physics is dominated by an annulus in momentum space with momenta $k \sim k_0$ and width 
$2\Lambda$. This implies that at high temperatures the model possess a
continuous symmetry in momentum space, or in other words, a large phase
space for fluctuations. The original model proposed by Brazovskii
contains also an
interaction term proportional to $\phi^4$. In the mean field
approximation, this model leads to a second order phase transition from
an isotropic phase at high temperatures to an anisotropic stripe phase with
modulation of the order parameter in the form:
\begin{equation}
\langle \phi(x) \rangle = A\ \cos{(k_0\,x)}.
\end{equation}
Working in three dimensions, Brazovskii showed
that including fluctuations of the order parameter self-consistently
leads to a ``fluctuation-induced first order transition''. Subsequently
this transition was observed and studied in diblock
copolymers~\cite{BaRoFrGl1988}. In two dimensions, stripe phases
arising from competing interactions are also observed in many systems,
a notable example being that of 2d ultrathin ferromagnetic films with
perpendicular anisotropy, in
which the short range exchange interaction between spins is frustrated
by the long range character of the dipolar interaction, giving rise to
the well known magnetic domains~\cite{HuSc1998}. In recent years
there have been indications that a mechanism similar to that proposed by Brazovskii
can be at work in these systems~\cite{CaStTa2004}. 
However,  stripe solutions
are not stable with rigorously isotropic interactions, and fluctuations
in the stripe direction diverge
logarithmically in 3d. Then, for two dimensional systems the situation
should be worse unless some isotropy-breaking effect be at work, like,
e.g. lattice effects~\cite{Le2007}. Nevertheless, even if positional
long range order is forbidden for such models in 2d, one can ask if
some kind of orientational order, reminiscent of stripe order, may
survive in an isotropic model of the kind considered.

Recently~\cite{BaSt2007}, we analyzed which kind of interaction terms could
give rise to {\em purely} orientational order in two dimensions, besides the
already known Brazovskii stripe solutions, which posses orientational
as well as translational long range order. Considering a generic
interaction term of the form:
\begin{widetext}
\begin{equation}
H_{\rm int}=\int_\Lambda \frac{d^2k_1}{(2\pi)^2}\frac{d^2k_2}{(2\pi)^2}\frac{d^2k_3}{(2\pi)^2}\frac{d^2k_4}{(2\pi)^2}   
\; u(\vec k_1,\vec k_2,\vec k_3,\vec k_4)\; \phi(\vec k_1)\phi(\vec k_2)\phi(\vec k_3)\phi(\vec k_4) 
\; \delta(\vec k_1+\vec k_2+\vec k_3+\vec k_4).
\label{H4}
\end{equation} 
\end{widetext}
and performing an analysis in the context of the Renormalization
Group, we found~\cite{BaSt2007} that {\em as a consequence of the isotropy of
  interactions in 2d} the space dependent function $u(\vec k_1,\vec k_2,\vec
k_3,\vec k_4)$ should depend solely on an angle $\theta$ and have the form:
\begin{equation}
u(\theta)=u_0+ u_2\;\cos(2\theta)+u_4\;\cos(4\theta)+\ldots
\label{fourier}
\end{equation}
Here, the first term $u_0$ leads to the usual $\phi^4$ theory
considered by Brazovskii in his model for the isotropic-stripe 
transition. The other terms are all relevant in the RG sense and then
the theory is not renormalizable. Nevertheless, it is evident that all
the terms share the nematic symmetry $\theta\to \theta+\pi$. 
We then
proceeded to analyze the effect of the first of those terms, proportional to
$\cos{(2\theta)}$. From the definitions in section \ref{op}, we realize that the
$\cos{(2\theta)}$ factor can be conveniently expressed in terms of the
tensor order parameter and the interaction energy can be written in the
form:
\begin{equation}
H_{\rm int}=\int d^2x\;\; \left\{  u_0\; \phi^4(\vec{x}) + u_2\; tr\
\hat Q ^2
+ \gamma\; tr\ \hat Q ^4\right\}
\label{HN}
\end{equation}
with $\gamma > 0$ and 
\begin{equation}
\hat Q_{ij}(\vec{x})= \phi(\vec{x}) \left(\nabla_i\nabla_j-\frac{1}{2}\nabla^2\delta_{ij}\right)\phi(\vec{x}).
\label{orderparameter}
\end{equation}
The gradients are related to the director $\hat n_i =
\nabla_i/|\nabla_i|$. From Eq.(\ref{orderparameter}) it is clear that
the nematic order parameter is essentially a quadrupolar moment.

Next, we proceed to analyze this extended model
in the self consistent Hartree approximation.

\subsection{Hartree approximation}

Replacing in eq. (\ref{HN}) 
$\phi^4\to \phi^2 <\phi^2>$ and 
$ tr\ \hat Q ^2\to tr\ \left\{ \phi \left(\nabla_i\nabla_j-\frac{1}{2}\nabla^2\delta_{ij}\right)\phi \right\}
\langle\hat Q_{ij}(\vec{x})\rangle$, where the mean values have to be determined self consistently,
we obtain a quadratic Hamiltonian in the Hartree approximation, which in
reciprocal space reads:
\begin{equation}
H_{\rm Hartree}=\frac{1}{2}\int \frac{d^2k}{(2\pi)^2}\,
\phi(\vec{k})\left(\beta^{-1}C^{-1}(\vec{k})\right)  \phi(-\vec{k}),
\end{equation}
with the two-point correlator $C(\vec{k})$ given by~\cite{BaSt2007}:
\begin{equation}
C(\vec{k})=\frac{T}{r + \frac{1}{m}(k-k_0)^2-\alpha^2 k^2\cos(2\theta)( u_2+\gamma \alpha^2) }.
\label{C}
\end{equation}
Here
\begin{equation}
r=r_0+  u_0\int \frac{d^2k}{(2\pi)^2}\;\; C(\vec{k})
\label{rint}
\end{equation}
and
\begin{equation}
\alpha= \frac{1}{2} \int \frac{d^2k}{(2\pi)^2}\;  k^2 \cos(2\theta)\; C(\vec{k})
\label{alphaint}
\end{equation}
where we have chosen $\langle Q_{ij}\rangle=2\alpha \left(n_i
\, n_j-\frac{1}{2}n^2\delta_{ij} \right)$.  $\theta$ is the angle
subtended by $\vec k$ with the director $\hat n$. 

Equations (\ref{C}), (\ref{rint}) and (\ref{alphaint}) must be solved
self-consistently. Its solution has been discussed in Ref.\ 
\onlinecite{BaSt2007}. The main result which comes out is that 
in the case of attractive quadrupole interactions, $u_2<0$, equation (\ref{alphaint}) has non 
trivial solutions for the nematic order parameter. 
Writing the equations in terms of adimensional parameters $r\to
\tilde r T$, $k_0\to \tilde k_0\sqrt{mT}$ 
and $r_0\to \tau= a(1-T_c/T)$, one finds out that
for high temperatures, $T>T_c$, the only possible solution is $\alpha=0$. 
Nevertheless, at $T = T_c$ a nematic phase emerges continuously with $\alpha\sim
c(T_c-T)^{1/2}$ and the critical temperature  $T_c= \frac{2}{(m \tilde
  k_0^{1/2})}(\frac{u_0}{u_2})^{1/4}$. 

A spontaneously broken continuous symmetry in a two dimensional system
with short range isotropic interactions is forbidden by the
Mermin-Wagner theorem~\cite{MeWa1966}. In these systems, fluctuations
of the order parameter typically diverge and the precise nature of the
divergence can say if order is lost exponentially fast or if it decays
more slowly giving rise to what is called
``quasi-long-range-order''. In order to analyze the effect of
fluctuations of the nematic order parameter near the transition, we write the free energy of the model in the Hartree approximation.
The partition function is
\begin{equation}
Z=\int {\cal D}\phi\; e^{-\beta H_{\rm Hartree}} = e^{-\beta F_{H}}.
\end{equation}
Integrating over $\phi$ one arrives at
\begin{equation}
F_{H}=\frac{1}{2\beta} {\rm Tr}\ln\; C^{-1}.
\end{equation}
In the limit $(k-k_0)^2/m<<r_c$ (that is very near the Hartree critical temperature), the free energy reads
\begin{equation}
F_{H}=\frac{1}{2\beta} {\rm Tr}\ln\left\{\beta\left(r_c+\frac{(k-k_0)^2}{2m}-\bar\alpha^2 k^2 u_2\cos(2\theta) \right)\right\}
\end{equation}
where $r_c$ and $\bar\alpha$ are the solutions of the self-consistent
Hartree equations (\ref{rint}) and (\ref{alphaint}) described in Ref.\ \onlinecite{BaSt2007}.

Since near the transition $\bar\alpha^2 k_0^2u_2/r_c<< 1$, we  can
expand the free energy
in the form:
\begin{eqnarray}
F_{H}&\approx& \frac{1}{2\beta} {\rm Tr}\ln\left\{1-\frac{\bar\alpha^2 k^2 u_2}{r_c}\cos(2\theta)\right\} \nonumber \\
&\sim& -\frac{u_2\bar\alpha^2 }{2\beta r_c}{\rm Tr}\left\{k^2\cos(2\theta)\right\}
-\frac{u^2_2\bar\alpha^4 }{4\beta r^2_c}{\rm Tr}\left\{k^4\cos^2(2\theta)\right\},
\nonumber \\  && 
\end{eqnarray}
where a $k$-independent term was absorbed in $F_{H}$.
The first term of the expansion is zero by symmetry (upon integration
over $\theta$), therefore at leading order the free energy reduces to
\begin{equation}
F_{H}=-\frac{u^2_2\bar\alpha^4}{4\beta r^2_c}{\rm Tr}\left\{k^4\cos^2(2(\theta-\varphi))\right\}
\label{FHartree}
\end{equation}
This expression represents the contribution of the anisotropic
(nematic) part in the
Hartree approximation, very near the transition into the nematic phase
($\bar\alpha\neq 0$). In the last expression we have introduced the angle $\varphi$
that is the reference from which we measure the angle $\theta$
(which is the integration variable). 
At this level $\varphi$ is an arbitrary constant, as it should be in a spontaneous symmetry breaking scenario. 
Next we will study smooth fluctuation of this field.

\subsection{Fluctuations}

In the same way we have done in the Landau-Ginzburg theory (and for the same reasons),  we consider angle fluctuations of the order parameter, 
$
Q(x)=\bar\alpha\; \exp{i 2 \varphi(x)}. 
$
Therefore,  the free energy now takes the form: 
\begin{equation}
F_{H}=-\frac{u^2_2\bar\alpha^4 }{4\beta r^2_c}{\rm Tr}\left\{k^4\cos^2 2(\theta-\varphi(x))\right\}.
\end{equation}
The difficulty with this expression is the evaluation of the trace,
since its argument is not diagonal neither in $k$ 
 nor in $x$ space. 
To evaluate it, we  make a coarse graining of configuration space, in
such a way that in a small region around 
a point $x_0$ we consider $\varphi$ essentially constant. Then, we
can  average over all points $x_0$ 
covering all configuration space. This coarse grained free energy can
be diagonalized in $k$ space
 and the trace can be easily evaluated.  Consider the following expansion for $\varphi(x)$, for a fixed point $x_0$:
\begin{eqnarray}
\varphi(x)&=&\varphi(x_0)+\vec\nabla\varphi(x_0)\cdot (\vec x-\vec
          x_0)+ \dots \nonumber \\
          &\approx& \varphi'(x_0)+ \vec\nabla\varphi(x_0)\cdot \vec x
\end{eqnarray}
 where the constant $\varphi'(x_0)=\varphi(x_0)-\vec\nabla\varphi(x_0)\cdot \vec x_0$.
 With this expansion we rewrite the cosine in the expression for the free energy:
 \begin{eqnarray}
 \cos 2(\theta-\varphi)&\approx& \cos 2(\theta-\varphi'_0-\vec\nabla\varphi(x_0)\cdot \vec x)
  \nonumber \\
&\approx&\cos 2\theta'+2   ( \vec\nabla\varphi(x_0)\cdot \vec x) \; \sin 2\theta'
 \end{eqnarray}
  where $\theta'=\theta-\varphi'_0$ and we have considered smooth fluctuations
 $|\vec\nabla\varphi(x_0)|<<1$.  
 Therefore, 
 \begin{eqnarray}
 \cos^2 2(\theta-\varphi) &\approx& \cos^2 2\theta' +4\cos 2\theta' \sin 2\theta'\;\vec\nabla\varphi(x_0)\cdot \vec x\nonumber \\
  &+&4   (\vec\nabla\varphi(x_0)\cdot \vec x)^2 \; \sin^2 2\theta'
 \label{crossterm}
 \end{eqnarray}
 The first term  contributes with an additive constant to the free
 energy, and then we will not consider it anymore. The second term is identically
 zero by symmetry considerations, as shown in the appendix. The
 relevant leading contribution 
to the fluctuations is the last one. 
Thus, let us consider the coarse grained free energy for smooth fluctuations:
\begin{equation}
F_{fl}=-\frac{\bar\alpha^4 u^2_2}{\beta r^2_c}\int \frac{d^2x_0}{V}\;{\rm Tr}\left\{ k^4(\vec\nabla\varphi(x_0)\cdot \vec x)^2 \; \sin^2 2\theta'\right\},
\end{equation}  
where $V$ is the volume of the system.
Using the representation $\vec x=i \vec\nabla_k $, we write the trace
in $k$ space in the form:
\begin{equation}
F_{fl}=\frac{\bar\alpha^4 u^2_2}{\beta r^2_c}\int d^2x_0\int
\frac{dk}{(2\pi)^2} k d\theta \; 
\sin^2 2\theta \ (\vec\nabla\varphi(x_0)\cdot  \vec\nabla_k)^2 k^4.
\end{equation}
The $k$ derivatives can be evaluated as
\begin{equation}
(\vec\nabla\varphi(x_0)\cdot  \vec\nabla_k)^2 k^4
= 4 |\vec\nabla\varphi_0|^2 k^2 \left\{1+2 \sin^2\theta\right\},
\end{equation}
where $\vec\nabla\varphi_0\cdot \vec k= |\vec\nabla\varphi_0||\vec k|
\sin\theta$ because $\vec\nabla\varphi_0$ is in the direction of the
fluctuations (perpendicular to the director)
and then the angle between  $\vec\nabla\varphi_0$ and $\vec k$ is $\pi/2-\theta$. 

We finally obtain
\begin{equation}
F_{\rm fl}=\frac{\bar\alpha^4 u^2_2\Gamma}{\beta r^2_c}\int d^2x_0\; |\vec\nabla\varphi(x_0)|^2
\end{equation}
where
\begin{eqnarray}
\Gamma&=&4\int_{k_0-\Lambda}^{k_0+\Lambda} \frac{dk}{(2\pi)^2}\;k^3 \int_0^{2\pi} d\theta\; \sin^2 (2\theta)\; \left\{ 1+2 \sin^2\theta \right\}\nonumber \\ 
&=&\frac{1}{\pi} k_0^3\Lambda 
\end{eqnarray}
to leading order in the cut-off $\Lambda\sim \sqrt{m r_0}= \sqrt{m a(T_c-T)}$. 

Remembering that the Hartree solution of the order parameter is $\bar
\alpha= c (T_c-T)^{1/2}$, 
we can write the free energy for the fluctuations as:
\begin{equation}
F_{\rm fl}=K(T)\int d^2x\; |\vec\nabla\varphi(x)|^2
\label{KT}
\end{equation}
where the elastic constant is given by:
\begin{equation}
K(T)= \kappa \left(1-\frac{T}{T_c}\right)^{5/2}.
\label{frank}
\end{equation}
The constant $\kappa=(8/\pi)(a^{1/2}c^4 \tilde{k}_0^{1/6}u_2^{5/4})/( m^{7/3}u_0^{7/12})$.
Expression (\ref{KT}) is equal to equation (\ref{deltaf}).
Then, Gaussian fluctuations of the nematic order parameter around the
mean field solution diverge logarithmically, and the nematic
phase does not have true long range order, but instead retains
quasi-long-range order with the well known Kosterlitz-Thouless phenomenology~\cite{KoTh1973}.
The conclusion is that fluctuations change the nature of the phase transition and, in
particular, the critical temperature departs from its mean field value
$T_c$. The isotropic-nematic phase transition in the present model
takes place at a temperature
$T_{KT}$. At this temperature, a
continuous phase transition mediated by unbinding of disclinations
happens with $T_{KT}=(\pi/8)K(T_{KT})$~\cite{ToNe1981}. This relation,
together with (\ref{frank}), allows to obtain the transition
temperature as:
\begin{equation}
T_{KT} = \left(
\frac{1}{1+\frac{8 T_c}{\pi\kappa}}
\right) \; T_c
\end{equation}

\section{Conclusions \label{Conclusions}}
Systems with competing interactions can develop complex ordered
phases , with characteristics different from the usual ferromagnetic long range
 order. For many systems, competition may lead to ground states with
 modulations in the order parameter. These broken symmetry states
 naturally show orientational order, and sometimes also positional
 one. While positional long range order is strongly suppressed in two
dimensions for systems with isotropic interactions and continuous
 symmetry, orientational order
 is more robust. 
We have studied a rather general model for competing interactions at
different scales, looking for  conditions for the existence of a purely
orientational phase at low temperatures. 

We have shown that, in the two dimensional Brazovskii model, the quartic interactions 
with higher derivatives of the order parameter  are all relevant terms in the Renormalization Group sense. 
All these terms can be arranged and interpreted as representing
multipole interactions. Among them, 
the quadrupole-quadrupole interaction is the first non trivial contribution. 
A mean field solution of the model gives rise an isotropic-nematic
phase transition.
The analysis of Gaussian fluctuations around the mean field solution
leads to a phase diagram similar to the 
one found  by Toner and Nelson in the context of defect mediated
melting in two dimensions  ~\cite{ToNe1981}. Positional order of
Brazovskii stripe solutions is destroyed by thermal
fluctuations, which are known to diverge linearly in two dimensions. However, orientational quasi-long ranged 
order is preserved in the nematic phase of the extended model. 
We have shown that there is a critical temperature $T_{KT}$ at which
orientational quasi-long-range order is destroyed. By analogy with the
XY model, one can conclude that the disordering of the nematic phase
takes place by means of 
a disclination unbinding mechanism, and the isotropic-nematic
phase transition 
is in  the Kosterlitz-Thouless
universality class.

The main difference of our approach with that of the
Toner-Nelson-Kosterlitz-Thouless is that our model allows 
for an analysis of both sides of the phase transition. This fact makes
possible to characterize the 
transition in terms of ``microscopic'' parameters, which describe the underlying competing interactions. 
Also, within the present formalism, it is possible to alternatively
interpret the nematic phase as a 
quadrupole condensation rather than a melting of topological defects.

Finally, the presence of a nematic phase from competing interactions in two
dimensions can be present in a variety of systems like ultrathin
ferromagnetic films with perpendicular anisotropy~\cite{PoVaPe2003}, 
block copolymers~\cite{SeHeHaKr1993},
 microemulsions and colloids~\cite{ZaWiMaSeNi2003,ImRe2006}, between others. The detection and
quantitative characterization of such phases in those systems rely on novel
imaging techniques which are at present rapidly evolving.

\section{Appendix}
We show here that the second term in eq. (\ref{crossterm}) gives zero contribution to the free energy.
The contribution of this term to the free energy is:
\begin{equation}
F_2\sim 4i\int d^2k \cos 2\theta' \sin 2\theta'\;\vec\nabla\varphi(x_0)\cdot \vec \nabla_k k^4
\label{F2}
\end{equation}
with
\begin{equation}
\vec\nabla\varphi(x_0)\cdot \vec \nabla_k k^4= 4 k^2 \vec\nabla\varphi(x_0)\cdot \vec k=k^3 |\vec\nabla\varphi(x_0)| \sin\theta.
\end{equation}
The last expression is due to the fact that if we measure $\theta$
from the director and consider that 
$\vec\nabla\varphi(x_0)$ is orthogonal to it, then the angle between $\vec\nabla\varphi(x_0)$ and $\vec k$ is $\pi/2-\theta$. Then the scalar product is written in terms of $\cos(\pi/2-\theta)=\sin\theta$. Introducing in Eq. (\ref{F2}) and remembering that $\theta'=\theta-\varphi'_0$, 
\begin{eqnarray}
F_2&\sim& 16i |\vec\nabla\varphi(x_0)|\times\nonumber \\ 
&\times& \int dk k^4 \int_0^{2\pi} d\theta \cos 2(\theta-\varphi'_0) \sin 2(\theta-\varphi'_0)\sin\theta.
\nonumber \\ &&
\end{eqnarray}
The angular integral is identically zero whatever the value of $\varphi'_0$, thus 
$F_2=0$ .

\vskip 0.5cm

We acknowledge Yan Levin for useful comments and the 
Abdus Salam International Centre for Theoretical Physics, where part of this work was done.  
The {\em ``Conselho Nacional de Desenvolvimento Cient\'{\i }fico e Tecnol\'{o}gico
CNPq-Brazil''} and the {\em ``Funda{\c{c}}{\~{a}}o de Amparo {\`{a}} Pesquisa do Estado
do Rio de Janeiro''} are acknowledged for financial support.  
%

\begin{thebibliography}{10}

\bibitem{SeAn1995}
M. Seul and D. Andelman, Science {\bf 267},  476  (1995).

\bibitem{FrKi1999}
E. Fradkin and S.~A. Kivelson, Phys. Rev. B {\bf 59},  8065  (1999).

\bibitem{KiFrEm1998}
S.~A. Kivelson, E. Fradkin, and V.~J. Emery, Nature {\bf 393},  550  (1998).

\bibitem{AbKaPoSa1995}
A. Abanov, V. Kalatsky, V.~L. Pokrovsky, and W.~M. Saslow, Phys. Rev. B {\bf
  51},  1023  (1995).

\bibitem{VaStMaPiPoPe2000}
A. Vaterlaus {\it et~al.}, Phys. Rev. Lett. {\bf 84},  2247  (2000).

\bibitem{FrHe1987}
G.~H. Fredrickson and E. Helfand, The Journal of Chemical Physics {\bf 87},
  697  (1987).

\bibitem{BaRoFr1990}
F.~S. Bates, J.~H. Rosedale, and G.~H. Fredrickson, The Journal of Chemical
  Physics {\bf 92},  6255  (1990).

\bibitem{ImRe2006}
A. Imperio and L. Reatto, The Journal of Chemical Physics {\bf 124},  164712
  (2006).

\bibitem{SeMoGoWo1991}
M. Seul, L.~R. Monar, L. O'Gorman, and R. Wolfe, Science {\bf 254},  1616
  (1991).

\bibitem{DiErGoJaLa1993}
A.~J. Dickstein {\it et~al.}, Science {\bf 261},  1012  (1993).

\bibitem{Br1975}
S.~A. Brazovskii, Sov. Phys. JETP {\bf 41},  85  (1975).

\bibitem{SwHo1977}
J. Swift and P.~C. Hohenberg, Physical Review A {\bf 15},  319  (1977).

\bibitem{HoSw1995}
P.~C. Hohenberg and J.~B. Swift, Phys. Rev. E {\bf 52},  1828  (1995).

\bibitem{BaSt2007}
D.~G. Barci and D.~A. Stariolo, Physical Review Letters {\bf 98},  200604
  (2007).

\bibitem{MeWa1966}
N.~D. Mermin and H. Wagner, Phys. Rev. Lett. {\bf 17},  1133  (1966).

\bibitem{KoTh1973}
J.~M. Kosterlitz and D.~J. Thouless, Journal of Physics C: Solid State Physics
  {\bf 6},  1181  (1973).

\bibitem{ToNe1981}
J. Toner and D.~R. Nelson, Phys. Rev. B {\bf 23},  316  (1981).

\bibitem{BaSt2007-2}
D.~G. Barci and D.~A. Stariolo, Physical Review Letters {\bf 99},  228904
  (2007).

\bibitem{Le2007}
Y. Levin, Physical Review Letters {\bf 99},  228903  (2007).

\bibitem{deGPr1998}
P.~G. de~Gennes and J. Prost,  in {\em The {P}hysics of {L}iquid {C}rystals},
  edited by O.~U. Press (Oxford University Press, Oxford, 1998).

\bibitem{BaRoFrGl1988}
F.~S. Bates, J.~H. Rosedale, G.~H. Fredrickson, and C.~J. Glinka, Phys. Rev.
  Lett. {\bf 61},  2229  (1988).

\bibitem{HuSc1998}
A. {Hubert} and R. {Schafer}, {\em Magnetic Domains} (Springer-Verlag, Berlin,
  1998).

\bibitem{CaStTa2004}
S.~A. Cannas, D.~A. Stariolo, and F.~A. Tamarit, Phys. Rev. B {\bf 69},  092409
   (2004).

\bibitem{PoVaPe2003}
O. Portmann, A. Vaterlaus, and D. Pescia, Nature {\bf 422},  701  (2003).

\bibitem{SeHeHaKr1993}
R.~A. Segalman, A. Hexemer, R.~C. Hayward, and E.~J. Kramer, Macromolecules
  {\bf 36},  3272  (1993).

\bibitem{ZaWiMaSeNi2003}
K. Zahn {\it et~al.}, Phys. Rev. Lett. {\bf 90},  155506  (2003).

\end{thebibliography}
%

%
\end{document}